\documentclass[nofootinbib, twocolumn, secnumarabic, amssymb, longbibliography, aps, prb]{revtex4-2}
\usepackage{graphicx}
\usepackage{bm}
\usepackage{dcolumn}

\usepackage{graphicx,wrapfig,lipsum}
\usepackage{moresize}
\usepackage{xcolor}

\begin{document}
\title{Muon spectroscopy investigation of anomalous dynamic magnetism in NiI$_2$}
\author{T.L. Breeze$^{1}$, B.M. Huddart$^{1,2}$, A. Hern\'andez-Meli\'an$^{1}$,  N.P. Bentley$^{1}$,
     D.A. Mayoh$^{3}$, G.D.A. Wood$^{3,4}$, G. Balakrishnan$^{3}$, J. Wilkinson$^{4}$, F.L. Pratt$^{4}$, T.J. Hicken$^{5}$, S.J. Clark$^{1}$, T. Lancaster$^{1}$.}
\affiliation{$^{1}$Department of Physics, Center for Materials Physics, Durham University, Durham,
DH1 3LE, United Kingdom\\
$^{2}$Clarendon Laboratory, University of Oxford, Department of Physics, Oxford OX1 3PU, United Kingdom\\
$^{3}$Department of Physics, University of Warwick, Coventry, CV4 7AL, United Kingdom\\
$^{4}$ISIS Facility, STFC-Rutherford Appleton Laboratory, Harwell Science and Innovation Campus, Didcot, OX11 0QX, United Kingdom\\
$^{5}$PSI Center for Neutron and Muon Sciences CNM, 5232 Villigen, Switzerland
}
\date{\today}

\begin{abstract}
  We present the results of muon-spin relaxation ($\mu^{+}$SR) measurements of the van der Waals magnet NiI$_2$, which probe  magnetic phase transitions at $T_{\mathrm{N1}}=73$~K and $T_{\mathrm{N2}}=60$~K. Supporting density functional theory (DFT) calculations allow the determination of a single muon stopping site whose magnetic environment is consistent with the proposed ground-state magnetic structure. $\mu^{+}$SR measurements of the dynamics reveal behavior consistent with spin-wave excitations below $T_{\mathrm{N2}}$. In the region $T_{\mathrm{N2}}<T<T_{\mathrm{N1}}$ the character of the dynamics changes qualitatively, resulting in an unusual region of temperature-independent fluctuations.
\end{abstract}

\maketitle

\section{Introduction}

Magnetic van der Waals materials are two-dimensional (2D) crystals 
containing magnetic elements, expected to
exhibit intrinsic low-dimensional magnetic properties. These cleavable materials
provide a platform for exploring magnetism in 
the 2D limit, where a range of emergent phenomena are expected \cite{burch2018magnetism,wang2022magnetic}. Aims in this field include the stabilization of
topological spin textures, such as vortices, skyrmions or merons \cite{lancaster2019skyrmions}, or the
realization of novel topological magnetic states of matter,
featuring topological order or bandstructures \cite{kezilebieke2020topological,zhang2022two}.
NiI$_{2}$ is a member of the
transition-metal dihalides, a family of materials previously noted for multiferroic behavior \cite{mcguire2017crystal}, with 
NiI$_2$ itself displaying ferroelectricity and antiferromagnetic order in its ground state \cite{kurumaji2013magnetoelectric}.
Recently, it has been proposed that it hosts a novel form of skyrmion phase in the 2D limit \cite{amoroso2020spontaneous,blei2021synthesis,song2022evidence}.
Muon spectroscopy is a sensitive probe of low-dimensional magnetism and the dynamics resulting from topological excitations \cite{lancaster2016transverse,amato2014understanding,hicken2020magnetism,hicken2021megahertz,hicken2022energy}. Here we use the technique to investigate the magnetic transitions in NiI$_{2}$ and the low-temperature magnetic dynamics. The magnetic ground state is shown to host conventional spin wave dynamic excitations, despite its complicated helical magnetic structure. The regime close to the magnetic ordering temperature is found to have a distinct and unconventional temperature-independent dynamic signature.

      \begin{figure}
\includegraphics[width=70mm]{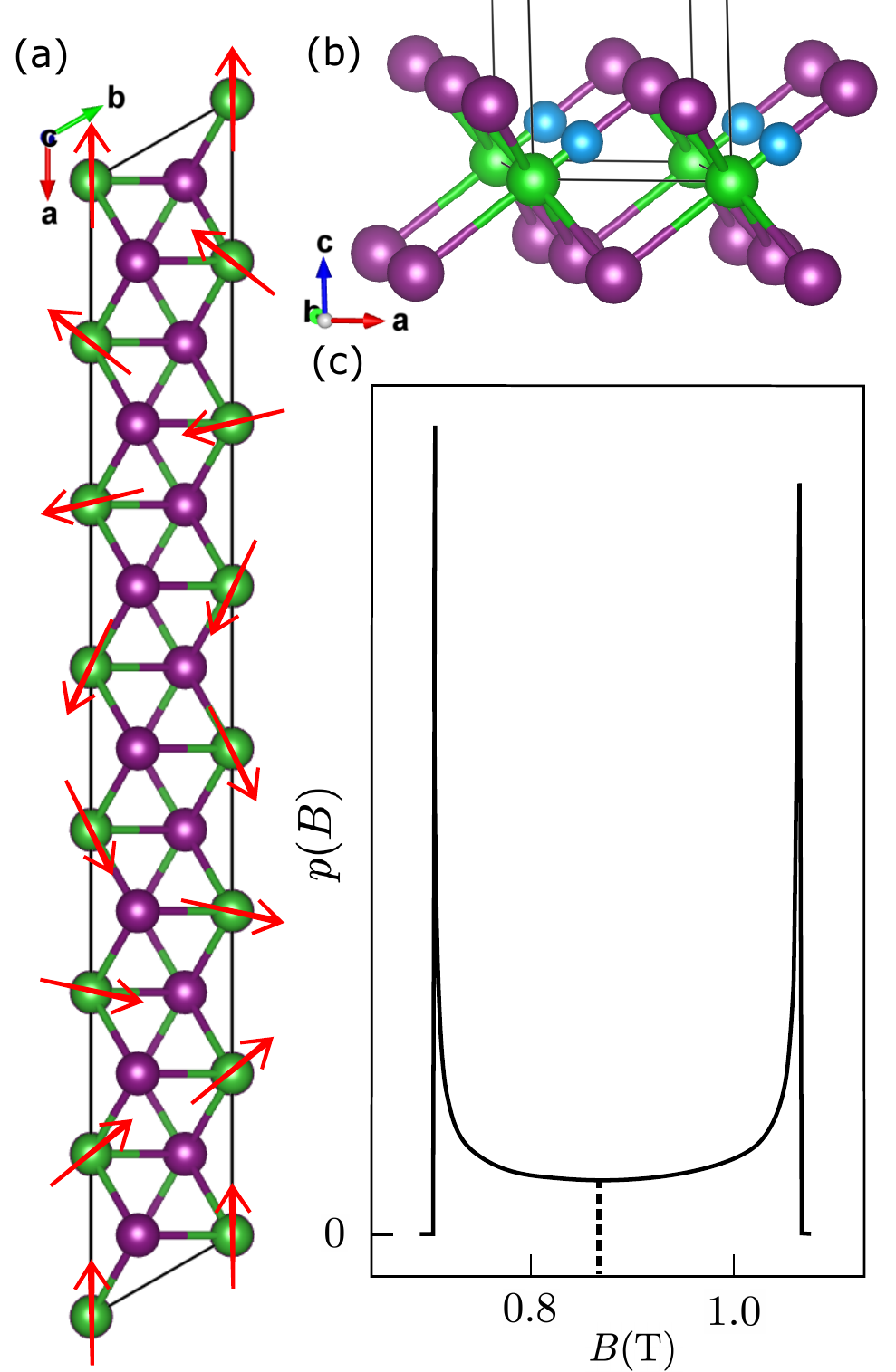}
       \vspace{0cm}
       \caption{
(a) NiI$_2$ structure and magnetic ground state showing Ni ions (green) and I ions (purple). The structure comprises stacked trigonal layers of magnetic Ni ions. Arrows indicate spins. (b) Low-energy candidate muon stopping site (blue) with (c) the associated field distribution, $p(B)$, (arbitrary units) with a mean field of 0.87~T indicated by the dashed line.
\label{fig:structure}
       }
\vspace{-0.5cm}
\end{figure}

NiI$_{2}$ is a centrosymmetric magnetic
semiconductor long known for its helimagnetism \cite{day1976optical,day1980incommensurate,billerey1977neutron,kapeghian2024effects,liu2020vapor,kurumaji2013magnetoelectric}.
The material crystallizes in the rhombohedral CdCl$_{2}$-type
structure ($R\bar{3}m$) 
with a magnetic Ni$^{2+}$ ion 
($S = 1$) carrying an ordered moment of 1.6$\mu_{\mathrm{B}}$ \cite{kurumaji2013magnetoelectric,mcguire2017crystal}. A
 single
layer of NiI$_{2}$ is characterized by a triangular net of magnetic
cations  and competing ferromagnetic and
antiferromagnetic interactions, resulting in strong magnetic
frustration.
 The static bulk magnetic susceptibility of single crystals show features at $T_{\mathrm{N}1}=76$~K and $T_{\mathrm{N}2}=58$~K that suggest two successive antiferromagnetic phase transitions in zero applied magnetic field \cite{kurumaji2013magnetoelectric}.
 Evidence of these transitions is also seen in features in the specific heat capacity \cite{billerey1977neutron}. The higher-temperature transition at $T_{\mathrm{N}1}$ takes the system from a paramagnetic high-temperature phase, to an antiferromagnetic phase on cooling in which,
for $T_{\mathrm{N}2}< T < T_{\mathrm{N}1}$, the magnetic order comprises ferromagnetic planes coupled antiferromagnetically along the $c$ axis \cite{kurumaji2013magnetoelectric}.
 At the lower-temperature transition at $T_{\mathrm{N}2}$ the symmetry of the crystal structure changes from trigonal to monoclinic with decreasing temperature, owing to a slight shift in the Ni layers along the $a$-direction giving an overall tilt. 
This structural distortion causes the spin texture to transform into a screw-helimagnetic ground state with propagation vector $\bold{q}=(0.138, 0, 0.1457)$ in the lattice basis \cite{kuindersma1981magnetic}.
Here the $\bold{q}$ vector is slanted from the triangular-lattice
basal plane and, correspondingly, the spin-spiral plane is also
canted from the plane that includes the $[001]$ axis (Fig~\ref{fig:structure}).
In the resulting helimagnetic
screw-spin ordered ground state,
NiI$_{2}$ shows spin-driven ferroelectricity, while the intermediate magnetic state between $T_{\mathrm{N}1}$ and $T_{\mathrm{N}2}$ is
paraelectric \cite{kurumaji2013magnetoelectric}. 


Magnetically, the transition at $T_{\mathrm{N}1} = 76$~K
has not been investigated in detail, with several earlier studies focusing on the low-temperature multiferroic properties \cite{song2022evidence,fumega2022microscopic,ju2021possible,lebedev2023electrical}. The prediction of skyrmionic spin excitations close to the ordering temperature in monolayers of NiI$_{2}$ motivates this investigation of the ordering behavior of the bulk material and its dynamics at a local level. 
We present the results of $\mu^+$SR experiments on high-quality single crystal and polycrystalline samples of NiI$_2$ along with discussion of the results paired with an analysis of candidate muon stopping sites and the dynamics in the ordered regime. Our results, while being consistent with the reported magnetic ground state of the system, also suggest that the temperature regime $T_{\mathrm{N2}}<T<T_{\mathrm{N1}}$ is characterized by an unusual spectrum of fluctuations in which the observed muon relaxation is temperature independent, whereas at temperatures below $T_{\mathrm{N2}}$ the material exhibits more conventional dynamic magnetism.

We grew single crystals of NiI$_2$ by the Bridgman method \cite{yoshinaga2004bulk}. Stoichiometric quantities of Ni and I were inserted into a quartz tube which was then sealed under vacuum. The tube was inserted into a vertical Bridgman furnace and slowly heated to 750$^\circ$C. The tube was then slowly cooled at 1$^\circ$C/hr to 600$^\circ$C. Once at 600$^\circ$C the tube was then rapidly cooled to room temperature and then removed from the furnace. We have characterized our samples by carrying out field-cooled magnetic susceptibility measurements, which are also in agreement with previous results \cite{kurumaji2013magnetoelectric}.


%

      \begin{figure} 
\begin{center}
\includegraphics[width=85mm]{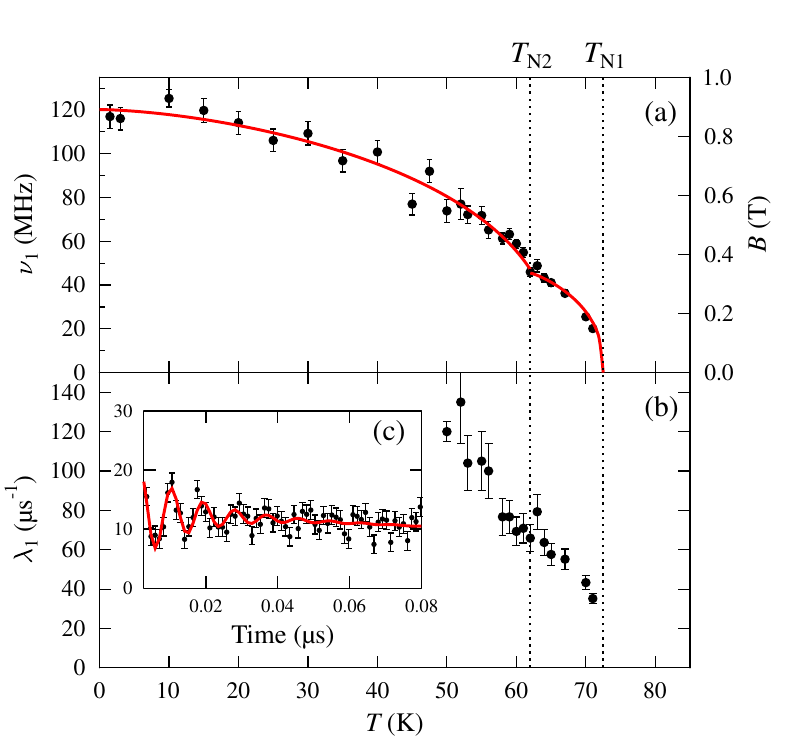}
       \vspace{0cm}
       \caption{
         Temperature evolution of ZF $\mu^{+}$SR fitting parameters. (a) The precession frequency $\nu_1$  with equivalent local field $B$ shown on right-hand axis and (b) the associated relaxation rate $\lambda_1$. Inset: example $\mu^+$SR spectrum at 3~K (c) with fit in red, plotted as percentage asymmetry.  The frequency  $\nu_1$ in (a) drops to zero at $T_{\mathrm{N}1}$, indicating the absence of long-range order above this temperature. The transition at $T_{\mathrm{N}2}$ is  seen in a discontinuous change in the gradient of $\nu$. Red lines are fits generated using Eqs~\ref{eqn:asym_fit_PSI} and \ref{eqn:freq_fit}. 
       In (b) the  relaxation rate $\lambda_1$ is fixed at 120~$\mu$s$^{-1}$ in the fitting routine below 50~K. 
\label{fig:fits}
       }
\end{center}
\vspace{-0.5cm}
\end{figure}

\section{Experimental methods}\label{sec2}

In a $\mu^{+}$SR experiment spin-polarized muons are implanted in a sample where they interact with the local magnetic field at the muon site.
After, on average, 2.2~$\mu$s, the muons decay into a positron and two neutrinos. By detecting the positrons, which are preferentially emitted in the direction of the muon spin at the time of decay, we can track the polarization of the muon-spin ensemble \cite{blundell2022muon}.
In a zero-field (ZF) $\mu^{+}$SR experiment the local magnetic field at the muon sites arises largely due to the configuration of the spins in the system.
When the muon-spin has a component perpendicular to the local field $B$, precession occurs with angular frequency $\omega=\gamma_\mu B$, where $\gamma_\mu=2\pi\times135.5$~MHz~T$^{-1}$ is the gyromagnetic ratio of the muon.
When the muon-spin aligns with the local field, only dynamic fluctuations can depolarize the muon-spin ensemble.
The quantity of interest in the experiment is the asymmetry $A\left(t\right)$, calculated from  counts measured in the detectors forwards and backwards of the initial muon-spin polarization direction $N_{\rm F,B}$, corrected using a parameter $\alpha$  reflecting detector efficiency, via $A\left(t\right)=\left(N_{\rm F}-\alpha N_{\rm B}\right)/\left(N_{\rm F}+\alpha N_{\rm B}\right)$.
The asymmetry is directly proportional to the polarization of the muon ensemble. Spectra were fitted using the WiMDA fitting program \cite{pratt2000wimda}.

\section{Results and discussion}

We made zero-field (ZF)  $\mu^{+}$SR measurements at temperatures from 1.7~K to 80~K using the GPS spectrometer at  S$\mu$S \cite{amato2017new}, with points concentrated close to the two transitions. Measurements were made on a single crystal sample consisting of a disc of NiI$_{2}$ with the $c$ axis parallel to the initial muon-spin polarization.
Below $T_{\mathrm{N1}}$ oscillations are observed in the muon asymmetry, corresponding to long-range magnetic order (LRO) occurring throughout the bulk of the material. 
We find that a good fit is achieved with only one oscillatory frequency with no phase offset, indicating a dominant contribution to the asymmetry signal from a single muon stopping site.
We fit the measured asymmetry to a model of the form
\begin{equation}\label{eqn:asym_fit_PSI}
A(t)=A_1{\rm e}^{-\lambda_1t}\cos(2\pi\nu_1t)+A_2{\rm e}^{-\lambda_2t}+A_3,
\end{equation}
where the first term corresponds to a relaxing precession of the muon spin components perpendicular to the local magnetic field. Terms two and three are relaxing and constant background components respectively, with the former due to those muon-spin components initially parallel to the internal magnetic field. For our single crystal measurement we find that the ratio $A_1:A_2$ is roughly 3:1. 
 Over the range of temperatures up to 72~K the data exhibit very fast damping of the oscillatory term (with the oscillations only visible for roughly the first 0.1~$\mu$s). We find the oscillating component relaxes much faster than the purely exponential component and the short time window makes changes in $\lambda_2$ difficult to fit. In fitting the data the relaxing amplitudes $A_1$ and $A_2$ are held constant, as is $\lambda_2$. Above $T\approx 72$~K the oscillations are no longer visible.
Fig~\ref{fig:fits} shows plots of the fitted parameters, including frequencies $\nu_1$ against temperature. The data above 62~K are fitted to a model
\begin{equation}\label{eqn:freq_fit}
\nu_1(T)=\nu_0[1-(T/T_{\rm N})^\delta]^{\beta}\text{,}
\end{equation}
where $T_{\rm{N}}$ is a critical temperature and $\nu_1$ is an effective order parameter for the magnetic phase transition. We find $T_{\mathrm{N1}}=72.5(1)$~K (Fig~\ref{fig:fits}) with exponents $\delta=1.20(1)$ and $\beta=0.39(1)$, fairly typical of a three-dimensional Heisenberg system.
At 62~K there is a subtle change in behavior of $\nu_1$ involving a discontinuous change in the  gradient of the measured frequency [Fig~\ref{fig:fits}(a)] corresponding to the transition at $T_{\mathrm{N2}}$.

We carried out supporting muon-site computations using the MuFinder software \cite{huddart2022mufinder}. We populated a supercell of the structure made up of $2\times 2\times 2$ conventional unit cells with muons at random positions with the constraint of a minimum distance between each initial muon position of 1~\AA \ and a minimum muon-atom distance of 1~\AA. Geometry optimizations were then performed using the {\sc Castep} \cite{clark2005first} code to relax the geometry of the full structure, causing muons to fall into local minima of energy. Calculations were done using the PBE functional \cite{perdew1996generalized}, with a plane wave basis set cut-off energy of 465~eV and a k-point grid of size $2\times 2\times 1$. Following a geometry optimization, muons were grouped into symmetry-equivalent positions so that they can be clustered into similar stopping sites. 
Our measurements of both the monoclinic ($T<T_{\mathrm{N2}}$) and trigonal ($T_{\mathrm{N2}}<T<T_{\mathrm{N1}}$) structures are consistent with a single low-energy stopping site. The minimum energy sites for both structural phases are found to share the same local environment, between the Ni ion and the nearest I$^{-}$ ion pictured  in Fig~\ref{fig:structure}(b). Using this candidate stopping site we can calculate the local dipole field experienced by implanted muons and then construct simulated $\mu^{+}$SR spectra for comparison with the data. 
The simulated polarization is computed using
\begin{equation}\label{eqn:spectrum}
P_z(t)=\sum_i f_{z,i}^2 p_i +\sum_{i}\left(1-f_{z,i}^2\right)p_i \cos(\gamma_{\mu} B_i t),
\end{equation}
where the probabilities $p_i$ correspond to the distribution of field strengths felt by muons stopping in the sample, $B_{i}$ is the magnitude of the local field and $f_{z,i}=B_{z,i}/B_i$ is the computed ratio of the $z$ component of magnetic field to the overall local field magnitude.

The low-temperature magnetic ground state is predicted to be helimagnetic with a spin spiral which has a component along the $a$ axis that results in one complete rotation of the spin across roughly seven unit cells \cite{fumega2022microscopic,kurumaji2013magnetoelectric}, described by the spin propagation vector $\mathbf{q}=(0.138, 0, 0.1457)$. Computing the resulting local dipole field with an ordered Ni ion moment of 1.6$\mu_{\mathrm{B}}$ \cite{mcguire2017crystal} at the predicted low-energy muon site yields a distribution of fields felt by muons in structurally equivalent but magnetically inequivalent sites [Fig~\ref{fig:structure}(c)]. 
This distribution shows two prominent peaks, and a simulated spectrum is therefore dominated by oscillations at two frequencies with associated fields corresponding to the peaks.
However, our muon site calculations indicate that there is a very shallow local minimum in energy about the candidate site, and a large number of sites with small displacements (all within 0.4~\AA \ of each other) from the average are found in the calculation. Each displacement will result in a field distribution similar to that shown in Fig~\ref{fig:structure}(c), but with the two peaks at different values of local field. Taking the normalized sum of the dipole field contributions from each of these sites therefore broadens the two peaks such that they overlap, resembling a single gaussian peak centred on an average field of 0.867~T, equivalent to a precession frequency of 117~MHz. This frequency is in excellent agreement with our single measured frequency of 117(5)~MHz close to $T=0$~K.
Finally, calculating a simulated zero temperature spectrum from this overall probability distribution and applying the same fitting function used for the data gives a crude estimate for the relaxation of $\lambda_1=175~\mu$s$^{-1}$. Compared to our experimental value of $\lambda_1=120(18)~\mu$s$^{-1}$, this suggests that the relaxation in the oscillating component is largely accounted for by decoherence due to this distribution of local fields, with relatively little dynamic contribution. Spin-DFT calculations in other work on this material suggests that the magnetic ground state is complicated by a magnetisation of the I nuclei by the Ni ions \cite{fumega2022microscopic}. We found, however, that we were able to produce simulated spectra that agreed well with experiment assuming magnetic centres on the Ni nuclei only.

In the intermediate temperature regime $T_{\mathrm{N2}}<T<T_{\mathrm{N1}}$, the magnetic structure is predicted to be a simple antiferromagnet with the spin propagation vector (0, 0, 1/2). A similar dipole field analysis in this case gives an average field of $B=0.695$~T, corresponding to a $T=0$~K precession frequency of $\nu=93$~MHz for this structure. An extrapolation of the fitted frequency data in this region with the order parameter fit described above suggests a $T=0$~K frequency of $\nu=92(5)$~MHz, in good agreement with this theoretical value, though it should be noted that this fit is not strongly constrained owing to the small number of data points in this region.

\begin{figure} 
\begin{center}
\includegraphics[width=85mm]{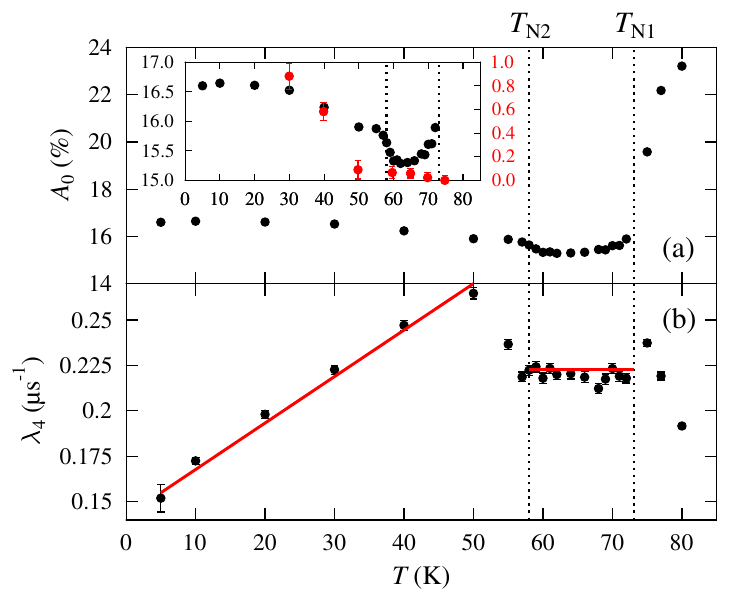}
       \vspace{0cm}
       \caption{Temperature evolution of parameters from fits to Eq~\ref{eqn:asym_fit_ISIS} for ZF data.
(a) Initial asymmetry, $A_0=A(t=0)$ (inset: $A_0$ in the region $T<85$~K [left axis, black] and fitted baseline, $A_8$, of the measurements in TF [right axis, red]) and (b) relaxation rate $\lambda_4$ with power law fits in different regions.
\label{fig:ISIS_data}
       }
\end{center}
\vspace{-0.5cm}
\end{figure}

We carried out further measurements on a polycrystalline sample using the HiFi instrument (at the STFC-ISIS Facility) \cite{king2013isis}, which allows for observation of the longer-time behavior in order to probe dynamics. ZF measurements were made across the same temperature interval as for the single-crystal measurements. In this case the oscillations are not resolvable as they occur over too short a period compared to the ISIS pulse width. Asymmetry spectra were therefore fitted to the function
\begin{equation}\label{eqn:asym_fit_ISIS}
A(t)=A_4{\rm e}^{-\lambda_4 t}+A_5,
\end{equation}
with a non-varying background term $A_5$ held constant at 13.6 (largely accounted for by muons stopping in the silver outside the sample), from which an initial asymmetry $A_0[=A(t=0)]$ can be calculated for each temperature via $A_0=A_4+A_5$. Results of the fitting procedure are shown in Fig~\ref{fig:ISIS_data}. We see indications of phase transitions at both $T_{\mathrm{N}1}$ and $T_{\mathrm{N}2}$, with $A_0$ decreasing from around 30~K, then undergoing a more rapid drop around $T_{\mathrm{N}2}\approx$~58~K followed by a sharp increase at $T_{\mathrm{N}1}\approx$~73~K.

We would expect the high temperature data ($T>T_{\mathrm{N}1}$) to include the full relaxing asymmetry $A_{\mathrm{\ssmall rel}}$ resulting from all muons implanted in the sample, excluding background contributions.
In a fully-ordered magnetic state in a powder we expect to see only 1/3 of the relaxing asymmetry resulting from muon spins initially parallel to the local field. The remaining 2/3, corresponding to spin-components initially perpendicular to the local field, result in oscillations outside of the ISIS frequency response window, that are not observed.
Notably, in the intermediate region $T_{\mathrm{N2}}<T<T_{\mathrm{N1}}$, an additional 1.4\% of asymmetry is lost, which implies that new relaxation channels are accessed in this regime that were not active below $T_{\mathrm{N2}}$. 

The dynamic signature of the magnetism is accessible via 
the longitudinal  relaxation rate $\lambda_{4}$, which rises approximately linearly up to a broad peak, which occurs  $\approx 10$~K below $T_{\mathrm{N}2}$. The rate $\lambda_{4}$ then falls to a constant value in the region between transitions, before increasing to a second peak at 73~K.
Such dynamics peaks are often indicative of local field fluctuation rates that decrease close to magnetic transitions. 
Below $T_{\mathrm{N}2}$ the evolution of $\lambda_{4}(T)$ is consistent with the fluctuations from spin-wave excitations in the case of a single gapless magnon band \cite{gomilsek2024anisotropic,beeman1968nuclear,janvsa2018observation}, for which we expect $\lambda\propto T^p$. We obtain a good fit to the data below 50~K with $p=1$, where
\begin{equation}\label{eqn:spin_wave}
p=\frac{2D}{s}-1.
\end{equation}
Here $D$ is the effective dimensionality (which we expect to be 1 here since the magnetic structure is described by a single propagation vector) and $s=1$ is the exponent for the spin wave dispersion relation $\omega=q^s$. Our exponent $p=1$ is consistent with $s=D=1$, corresponding to the helimagnetic magnetic structure predicted. This implies that the dynamics in this range are well described by contributions from a single magnon band that dominates the magnetic excitations.

In the  intermediate  temperature region $T_{\mathrm{N2}}<T<T_{\mathrm{N1}}$, we again expect $s=1$, owing to the simple antiferromagnetic spin texture. However, here the $T$-independence of the relaxation rate data suggests $p=0$, which cannot be satisfied for integer $D$. This suggests that above 60~K the dynamics we detect through the relaxation rate $\lambda_{4}$ are qualitatively different from those for $T<T_{\mathrm{N2}}$, and can no longer be described by a single gapless magnon band. Moreover, temperature dependent fluctuations suggest that the relaxation in this regime is no longer affected by the population of collective magnon excitations, nor dominated by the excitation of the system over an energy gap (e.g.\ that resulting from single-ion anisotropy arising from the spin-orbit interaction).
It is also possible that paraelectric fluctuations in this region complicate the observed dynamics, but the nature of the dynamics in this range are not made clear by our muon spectroscopy measurements.

It seems that the new magnetic structure stabilized when the crystal distorts above $T_{\mathrm{N2}}$
is such that the main relaxation channel for the muons ceases to be the result of magnons, although it is by no means obvious which excitations are dominant in this regime. We suggest however, that the changes in relaxation are ultimately driven by the distortion of the crystal. 
Specifically, below $T_{\mathrm{N2}}$ the local environment around the Ni nuclei changes symmetry as the unit cell goes from trigonal to monoclinic with decreasing temperature. Above $T_{\mathrm{N2}}$ the structure has the point group $D_{3d}$
at the Ni$^{2+}$ ion center and the partially filled 8-electron 3d shell of the Ni$^{2+}$ ion is split into three groups, with the d$_{x^2-y^2}$ and  d$_{xy}$, and the d$_{xz}$ and d$_{yz}$ orbitals forming two distinct degenerate bands separated in energy,  and the d$_{z^2}$ orbital occupying a different energy still. An observation of the shapes of the superpositions of these orbitals suggests that d$_{z^2}$ is likely to be the lowest energy state as it allows for electrons to be localized further away from the I$^{-}$ anions. This energy level structure will result in a partial filling of whichever is the higher energy out of the d$_{x^2-y^2}$ $+$ d$_{xy}$ and d$_{xz}$ $+$ d$_{yz}$ bands.
Below $T_{\mathrm{N2}}$ the structure becomes much less symmetric, losing centrosymmetry and allowing for the presence of the Dzyaloshinskii-Moriya (DM) interaction which results in the spin-canted non-collinear magnetic ground state in this phase. This lowering of symmetry also results in the loss of any rotation axis, meaning the structure must now take on one of three low-symmetry point groups at the Ni$^{2+}$ ion center, all of which would lift each of the orbital degeneracies, creating energy gaps between different electronic states in the 3d orbitals. It seems reasonable to suggest that this change in orbital occupancy and the possibility of a sizeable DM interaction leads to the altered magnetic structure with a qualitatively different spectrum of excitations.


In addition to our ZF measurements we made measurements in a weak applied transverse field (TF) for both single crystal (5~mT) and powder samples (2~mT). Weak TF measurements allow us to  determine the non-magnetic fraction of the material
as the muon spins in such regions
 precess in the small applied field, resulting in an oscillating component in the asymmetry with amplitude proportional to the nonmagnetic volume measured.
Muons stopping in ordered regions where the static field is much larger will not contribute to this oscillatory component.
We found that spectra for single crystals and powder were qualitatively similar and that parameters from fits to the two data sets behaved in the same way. After fitting the $\alpha$ parameter discussed in Section~\ref{sec2} from measurements made at temperatures above 70~K,
we fit the data to an equation of the form
\begin{equation}\label{eqn:asym_fit_TF}
A(t)=A_6{\rm e}^{-\lambda_6t}\cos(2\pi\nu_6t)+A_7{\rm e}^{-\lambda_7t}+A_8.
\end{equation}
Here the first term corresponds to precession of the muon spins in the nonmagnetic regions (which will include regions outside the sample) with a fixed frequency corresponding to the strength of the weak applied field, the second term is a relaxing exponential term, found to have a larger relaxation than that seen in ZF, and the third term is a constant baseline asymmetry that we allowed to vary in our fitting. We found that the amplitude $A_6$ remains constant up to 70~K, before increasing upon entering the paramagnetic region as expected. This implies that there is no change in the nonmagnetic fraction between the two low-$T$ ordered phases in the material, and therefore no macroscopic phase separation occurring in the material within the ordered regimes.

The fitted baseline asymmetry, $A_8$, for the powder data is plotted in the inset in Fig~\ref{fig:ISIS_data}. We observe a nonzero baseline asymmetry at low temperatures, which decreases on warming from 30~K to 50~K, above which it remains approximately  zero. 
A nonzero value of this term indicates the presence of muon spins that are not relaxed within the time window of the measurement. We would expect this in a conventional quasistatic magnetically ordered phase, where some muon spins will initially be parallel to the local field and will therefore be locked in direction during the measurement and not contribute to the relaxation.
The decrease in this baseline on warming coincides with our observation of a decrease in $A_{0}$ above 30~K in ZF, seen even before the system enters the
region of unconventional dynamics at temperatures $T_{\mathrm{N}2}<T<T_{\mathrm{N}1}$ on warming.
This suggests that at temperatures below $T_{\mathrm{N}2}$, dynamic relaxation channels are causing muon-spin flips, resulting in muon spins being relatively rapidly relaxed on the timescale of the ISIS frequency response. The baseline contribution vanishes at those temperatures above which the ZF relaxation rate $\lambda_{4}$ ceases to follow a linear temperature dependence. This is consistent with the new relaxation channels being activated as the system is warmed towards the region of more more complicated dynamics that characterizes the phase  at $T_{\mathrm{N}2}<T<T_{\mathrm{N}1}$.

The additional exponential term with amplitude $A_{7}$ was unexpected, but indicates the presence of an additional contribution to the relaxation, possibly from magnetic defects. This term is present in both the powder and the single crystal measurements and in both cases the amplitude $A_{7}$ and the relaxation rate $\lambda_{7}$ are constant on warming up to 50~K, and increase through the intermediate phase $T_{\mathrm{N}2}<T<T_{\mathrm{N}1}$. The term persists at temperatures just above the transition into disorder. We are not able to unambiguously discern what causes this term, but it may be explained by the presence of magnetic defects such as spins being frozen around stacking faults. These kinds of defects are likely to appear in layered materials such as NiI$_2$, and could be exacerbated in this case by the structural transition at $T_{\mathrm{N}2}$, about which there may be coexisting regions of both structural phases.


\section{conclusions}

In conclusion, muon-spin spectroscopy measurements have shown features corresponding to two successive phase transitions at $T_{\mathrm{N}1}\approx$~73~K and $T_{\mathrm{N2}}\approx$~60~K.
With the addition of a DFT muon stopping site analysis we have been able to identify a muon site located between neighbouring Ni and I ions, at which the muon spin behavior is consistent with the proposed magnetic structure, assuming we allow a small variation in the location where the site is realized.
The dynamic relaxation shows a linear increase in the region below 58~K, followed by a constant region between transitions, indicative of a qualitative change in excitation spectrum, which coincides with the change in magnetic structure.
It is worth noting that the prediction of skyrmions in a monolayer of this material was for a phase occurring at temperatures close to the ordering temperature, where skyrmions were presumably stabilized by thermal fluctuations. Given that we have identified two regimes of dynamics, it will be interesting in the future to see whether either region is able to support the equilibrium occurrence of skyrmion excitations in applied magnetic field in samples formed from a limited number of layers.

\section{Acknowledgements}

Muon measurements were made at the Swiss Muon Source and the STFC-ISIS Facility and we are grateful for the provision of beamtime. We also thank Chennan Wang and Hubertus Luetkens (S$\mu$S) for their assistance over the course of these measurements. Computational work was done using the Hamilton8 HPC service of Durham University and the ARCHER2 UK National Supercomputing Service (https://www.archer2.ac.uk). We acknowledge support from EPSRC (UK) (grant numbers: EP/N032128/1, EP/X035891/1 and EP/T005963/1 and T.L.B's studentship). N.P.B is grateful for the support of the Durham Doctoral Scholarship. Data will be made available via DOI:XXXXXXXXX.

\bibliography{NiI2_arxiv_1}

\begin{thebibliography}{10}

\bibitem{burch2018magnetism}
K.S. Burch, D.~Mandrus, and J.~Park.
\newblock {Magnetism in two-dimensional van der Waals materials}.
\newblock Nature \textbf{563}, 47 (2018).

\bibitem{wang2022magnetic}
Q.H. Wang, A.~Bedoya-Pinto, M.~Blei, A.H. Dismukes, A.~Hamo, S.~Jenkins,
  M.~Koperski, Y.~Liu, Q.~Sun, E.J. Telford, et~al.
\newblock {The magnetic genome of two-dimensional van der Waals materials}.
\newblock ACS nano \textbf{16}, 6960 (2022).

\bibitem{lancaster2019skyrmions}
T.~Lancaster.
\newblock Skyrmions in magnetic materials.
\newblock Contemp. Physics \textbf{60}, 246 (2019).

\bibitem{kezilebieke2020topological}
S.~Kezilebieke, M.N. Huda, V.~Va{\v{n}}o, M.~Aapro, Somesh~C. Ganguli, O.J.
  Silveira, S.~G{\l}odzik, A.S. Foster, T.~Ojanen, and P.~Liljeroth.
\newblock {Topological superconductivity in a van der Waals heterostructure}.
\newblock Nature \textbf{588}, 424 (2020).

\bibitem{zhang2022two}
G.~Zhang, H.~Wu, L.~Zhang, L.~Yang, Y.~Xie, F.~Guo, H.~Li, B.~Tao, G.~Wang,
  W.~Zhang, et~al.
\newblock {Two-Dimensional Van Der Waals Topological Materials: Preparation,
  Properties, and Device Applications}.
\newblock Small \textbf{18}, 2204380 (2022).

\bibitem{mcguire2017crystal}
M.A. McGuire.
\newblock Crystal and magnetic structures in layered, transition metal
  dihalides and trihalides.
\newblock Crystals \textbf{7}, 121 (2017).

\bibitem{kurumaji2013magnetoelectric}
T.~Kurumaji, S.~Seki, S.~Ishiwata, H.~Murakawa, Y.~Kaneko, and Y.~Tokura.
\newblock {Magnetoelectric responses induced by domain rearrangement and spin
  structural change in triangular-lattice helimagnets NiI$_2$ and CoI$_2$}.
\newblock Phys. Rev. B \textbf{87}, 014429 (2013).

\bibitem{amoroso2020spontaneous}
D.~Amoroso, P.~Barone, and S.~Picozzi.
\newblock {Spontaneous skyrmionic lattice from anisotropic symmetric exchange
  in a Ni-halide monolayer}.
\newblock Nat. Commun. \textbf{11}, 5784 (2020).

\bibitem{blei2021synthesis}
M.~Blei, J.L. Lado, Q.~Song, D.~Dey, O.~Erten, V.~Pardo, R.~Comin, S.~Tongay,
  and A.S. Botana.
\newblock {Synthesis, engineering, and theory of 2D van der Waals magnets}.
\newblock Appl. Phys. Rev. \textbf{8}, 021301 (2021).

\bibitem{song2022evidence}
Q.~Song, C.A. Occhialini, E.~Erge{\c{c}}en, B.~Ilyas, D.~Amoroso, P.~Barone,
  J.~Kapeghian, K.~Watanabe, T.~Taniguchi, A.S. Botana, et~al.
\newblock {Evidence for a single-layer van der Waals multiferroic}.
\newblock Nature \textbf{602}, 601 (2022).

\bibitem{lancaster2016transverse}
T.~Lancaster, F.~Xiao, Z.~Salman, I.O. Thomas, S.J. Blundell, F.L. Pratt, S.J.
  Clark, T.~Prokscha, A.~Suter, S.L. Zhang, et~al.
\newblock {Transverse field muon-spin rotation measurement of the topological
  anomaly in a thin film of MnSi}.
\newblock Phys. Rev. B \textbf{93}, 140412 (2016).

\bibitem{amato2014understanding}
A.~Amato, P.~Dalmas~de R{\'e}otier, D.~Andreica, A.~Yaouanc, A.~Suter,
  G.~Lapertot, I.M. Pop, E.~Morenzoni, P.~Bonf{\`a}, F.~Bernardini, et~al.
\newblock {Understanding the $\mu$ SR spectra of MnSi without magnetic
  polarons}.
\newblock Phys. Rev. B \textbf{89}, 184425 (2014).

\bibitem{hicken2020magnetism}
T.J. Hicken, S.J.R. Holt, K.J.A. Franke, Z.~Hawkhead,
  A.~{\v{S}}tefan{\v{c}}i{\v{c}}, M.N. Wilson, M.~Gomil{\v{s}}ek, B.M. Huddart,
  S.J. Clark, M.R. Lees, et~al.
\newblock {Magnetism and N{\'e}el skyrmion dynamics in GaV$_4$S$_{8-y}$Se$_y$}.
\newblock Phys. Rev. Research \textbf{2}, 032001 (2020).

\bibitem{hicken2021megahertz}
T.J. Hicken, M.N. Wilson, K.J.A. Franke, B.M. Huddart, Z.~Hawkhead,
  M.~Gomil{\v{s}}ek, S.J. Clark, F.L. Pratt, A.~{\v{S}}tefan{\v{c}}i{\v{c}},
  A.E. Hall, et~al.
\newblock Megahertz dynamics in skyrmion systems probed with muon-spin
  relaxation.
\newblock Phys. Rev. B \textbf{103}, 024428 (2021).

\bibitem{hicken2022energy}
T.J. Hicken, Z.~Hawkhead, M.N. Wilson, B.M. Huddart, A.E. Hall,
  G.~Balakrishnan, C.~Wang, F.L. Pratt, S.J. Clark, and T.~Lancaster.
\newblock {Energy-gap driven low-temperature magnetic and transport properties
  in Cr$_{1/3}$MS$_2$ (M=Nb,Ta)}.
\newblock Phys. Rev. B \textbf{105}, L060407 (2022).

\bibitem{day1976optical}
P.~Day, A.~Dinsdale, E.R. Krausz, and D.J. Robbins.
\newblock {Optical and neutron diffraction study of the magnetic phase diagram
  of NiBr$_2$}.
\newblock Journal of Phys. C \textbf{9}, 2481 (1976).

\bibitem{day1980incommensurate}
P.~Day and K.R.A. Ziebeck.
\newblock {Incommensurate spin structure in the low-temperature magnetic phase
  of NiBr$_2$}.
\newblock Journal of Phys. C \textbf{13}, L523 (1980).

\bibitem{billerey1977neutron}
D.~Billerey, C.~Terrier, N.~Ciret, and J.~Kleinclauss.
\newblock {Neutron diffraction study and specific heat of antiferromagnetic
  NiI$_2$}.
\newblock Phys. Lett. A \textbf{61}, 138 (1977).

\bibitem{kapeghian2024effects}
J.~Kapeghian, D.~Amoroso, C.A. Occhialini, L.G.P. Martins, Q.~Song, J.S. Smith,
  J.J. Sanchez, J.~Kong, R.~Comin, P.~Barone, et~al.
\newblock {Effects of pressure on the electronic and magnetic properties of
  bulk NiI$_2$}.
\newblock Phys. Rev. B \textbf{109}, 014403 (2024).

\bibitem{liu2020vapor}
H.~Liu, X.~Wang, J.~Wu, Y.~Chen, J.~Wan, R.~Wen, J.~Yang, Y.~Liu, Z.~Song, and
  L.~Xie.
\newblock {Vapor deposition of magnetic van der Waals NiI$_2$ crystals}.
\newblock ACS nano \textbf{14}, 10544 (2020).

\bibitem{kuindersma1981magnetic}
S.R. Kuindersma, J.P. Sanchez, and C.~Haas.
\newblock {Magnetic and structural investigations on NiI$_2$ and CoI$_2$}.
\newblock Physica B + C \textbf{111}, 231 (1981).

\bibitem{fumega2022microscopic}
A.O. Fumega and J.L. Lado.
\newblock {Microscopic origin of multiferroic order in monolayer NiI$_2$}.
\newblock 2D Mater. \textbf{9}, 025010 (2022).

\bibitem{ju2021possible}
H.~Ju, Y.~Lee, K.~Kim, I.H. Choi, C.J Roh, S.~Son, P.~Park, J.H. Kim, T.S.
  Jung, J.H. Kim, et~al.
\newblock {Possible persistence of multiferroic order down to bilayer limit of
  van der Waals material NiI$_2$}.
\newblock Nano letters \textbf{21}, 5126 (2021).

\bibitem{lebedev2023electrical}
D.~Lebedev, J.T. Gish, E.S. Garvey, T.K. Stanev, J.~Choi, L.~Georgopoulos, T.W.
  Song, H.Y Park, K.~Watanabe, T.~Taniguchi, et~al.
\newblock {Electrical Interrogation of Thickness-Dependent Multiferroic Phase
  Transitions in the 2D Antiferromagnetic Semiconductor NiI$_2$}.
\newblock Adv. Func. Mater. \textbf{33}, 2212568 (2023).

\bibitem{yoshinaga2004bulk}
M.~Yoshinaga, T.~Iida, M.~Noda, T.~Endo, and Y.~Takanashi.
\newblock {Bulk crystal growth of Mg$_2$Si by the vertical Bridgman method}.
\newblock Thin Solid Films \textbf{461}, 86 (2004).

\bibitem{blundell2022muon}
S.~Blundell, R.~De~Renzi, T.~Lancaster, and F.L. Pratt.
\newblock {\em {Muon Spectroscopy: An Introduction}}.
\newblock Oxford University Press, (2022).

\bibitem{pratt2000wimda}
F.L. Pratt.
\newblock {WIMDA: a muon data analysis program for the Windows PC}.
\newblock Physica B \textbf{289}, 710 (2000).

\bibitem{amato2017new}
A.~Amato, H.~Luetkens, K.~Sedlak, A.~Stoykov, R.~Scheuermann, M.~Elender,
  A.~Raselli, and D.~Graf.
\newblock {The new versatile general purpose surface-muon instrument (GPS)
  based on silicon photomultipliers for $\mu$SR measurements on a
  continuous-wave beam}.
\newblock Review of Scientific Instruments \textbf{88}, (2017).

\bibitem{huddart2022mufinder}
B.M. Huddart, A.~Hern{\'a}ndez-Meli{\'a}n, T.J. Hicken, M.~Gomil{\v{s}}ek,
  Z.~Hawkhead, S.J. Clark, F.L. Pratt, and T.~Lancaster.
\newblock {MuFinder: A program to determine and analyse muon stopping sites}.
\newblock Comp. Phys. Commun. \textbf{280}, 108488 (2022).

\bibitem{clark2005first}
S.J. Clark, M.D. Segall, C.J. Pickard, P.J. Hasnip, M.I.J. Probert, K.~Refson,
  and M.C. Payne.
\newblock {First principles methods using CASTEP}.
\newblock Zeitschrift f{\"u}r kristallographie-crystalline materials
  \textbf{220}, 567 (2005).

\bibitem{perdew1996generalized}
J.P. Perdew, K.~Burke, and Y.~Wang.
\newblock Generalized gradient approximation for the exchange-correlation hole
  of a many-electron system.
\newblock Phys. rev. B \textbf{54}, 16533 (1996).

\bibitem{king2013isis}
P.J.C. King, R.~de~Renzi, S.P. Cottrell, A.D. Hillier, and S.F.J. Cox.
\newblock {ISIS muons for materials and molecular science studies}.
\newblock Physica Scripta \textbf{88}, 068502 (2013).

\bibitem{gomilsek2024anisotropic}
M.~Gomilšek, T.J. Hicken, M.N. Wilson, K.J.A. Franke, B.M. Huddart,
  A.~Štefančič, S.J.R. Holt, G.~Balakrishnan, D.A. Mayoh, M.T. Birch, S.H.
  Moody, H.~Luetkens, Z.~Guguchia, M.T.F. Telling, P.J. Baker, S.J. Clark, and
  T.~Lancaster.
\newblock {Skyrmion and incommensurate spin dynamics in centrosymmetric Gd$_2
  $PdSi$_3$}.
\newblock arXiv preprint arXiv:2312.17323 (2024).

\bibitem{beeman1968nuclear}
D.~Beeman and P.~Pincus.
\newblock Nuclear spin-lattice relaxation in magnetic insulators.
\newblock Phys. Rev. \textbf{166}, 359 (1968).

\bibitem{janvsa2018observation}
N.~Jan{\v{s}}a, A.~Zorko, M.~Gomil{\v{s}}ek, M.~Pregelj, K.W. Kr{\"a}mer,
  D.~Biner, A.~Biffin, C.~R{\"u}egg, and M.~Klanj{\v{s}}ek.
\newblock {Observation of two types of fractional excitation in the Kitaev
  honeycomb magnet}.
\newblock Nature physics \textbf{14}, 786 (2018).

\end{thebibliography}
\bibliographystyle{unsrt}

\end{document}


\title{Supplemental Material: Muon spectroscopy investigation of anomalous dynamic magnetism in NiI$_2$}
\author{T.L. Breeze$^{1}$, B.M. Huddart$^{1,2}$, A. Hern\'andez-Meli\'an$^{1}$,  N.P. Bentley$^{1}$,
     D.A. Mayoh$^{3}$, G.D.A. Wood$^{3,4}$, G. Balakrishnan$^{3}$, J. Wilkinson$^{4}$, F.L. Pratt$^{4}$, T.J. Hicken$^{5}$, S.J. Clark$^{1}$, T. Lancaster$^{1}$.}
\affiliation{$^{1}$Department of Physics, Center for Materials Physics, Durham University, Durham,
DH1 3LE, United Kingdom\\
$^{2}$Clarendon Laboratory, University of Oxford, Department of Physics, Oxford OX1 3PU, United Kingdom\\
$^{3}$Department of Physics, University of Warwick, Coventry, CV4 7AL, United Kingdom\\
$^{4}$ISIS Facility, STFC-Rutherford Appleton Laboratory, Harwell Science and Innovation Campus, Didcot, OX11 0QX, United Kingdom\\
$^{5}$PSI Center for Neutron and Muon Sciences CNM, 5232 Villigen, Switzerland
}
\date{\today}
\maketitle

\section{Weak Transverse Field $\mu$SR}

      \begin{figure}
\begin{center}
\includegraphics[width=85mm]{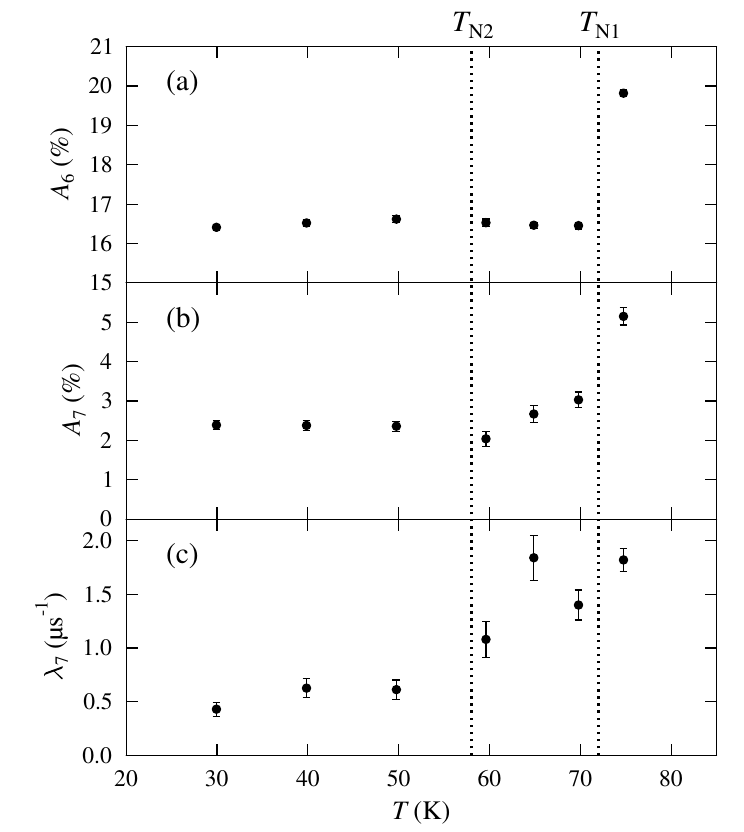}
       \vspace{0cm}
       \caption{Fitted parameters for our wTF data measured at ISIS. (a) The amplitude $A_6$, (b) the amplitude $A_7$ and (c) its associated relaxation rate $\lambda_7$.
\label{fig:TF}
       }
\end{center}
\vspace{-0.5cm}
\end{figure}

Fig~\ref{fig:TF} shows the fitted parameters $A_6$, $A_7$ and $\lambda_7$ from the fitting of the weak transverse field (wTF) data measured at ISIS to an equation of the form
\begin{equation}\label{eqn:asym_fit_TF}
A(t)=A_6{\rm e}^{-\lambda_6t}\cos(2\pi\nu_6t)+A_7{\rm e}^{-\lambda_7t}+A_8,
\end{equation}
\noindent (Equation 6 in the main text). We observe that the amplitude of the component corresponding to muon spin precession under an applied field, $A_6$, which is proportional to the magnetically disordered volume of the sample, is constant up to $T_{\mathrm{N}1}$, above which the sample is fully disordered. This indicates that there is no change in magnetic volume fraction across the structural transition at $T_{\mathrm{N}2}$. The presence of a significant non-zero value of $A_6$ below $T_{\mathrm{N}1}$ does not necessarily indicate a large disordered volume fraction at these temperatures, as much of this term will be accounted for by background effects such as muons stopping outside the sample in the silver backing plate. The terms $A_7$ and $\lambda_7$, corresponding to an unexpected term that is potentially explained by magnetic defects, are both constant up to $T_{\mathrm{N}2}$, and increase through the intermediate phase $T_{\mathrm{N}2}<T<T_{\mathrm{N}1}$. A single measurement taken at 135~K shows that this term vanishes at temperatures much higher than the transition to disorder.
